# Toward Eurasian SubMillimeter Telescopes: the concept of multicolor subTHz MKID-array demo camera MUSICAM and its instrumental testing

R.Duan[1], V.Khaikin[2], M.Lebedev[2], V.Shmagin[3], G.Yakopov[4], V.Vdovin[5], G.Bubnov[5], X. Zhang[1], C. Niu[1], S. Li[1,7], Di Li[1,6] and I. Zinchenko[5]

[1]National Astronomical Observatory, Chinese Academy of Science
[2]Special Astrophysical Observatory, Russia Academy of Science
[3]Institute of Astronomy, Russia Academy of Science
[4]Institute for Physical Problems, Russia Academy of Science
[5]Institute of Applied Physics, Russia Academy of Science
[6]NAOC-UKZN Computational Astrophysics Centre, University of KwaZulu-Natal, South Africa
[7]James Clerk Maxwell Telescope

**Abstract**

New challenges in submillimeter wave astronomy require instruments with a combination of high sensitivity andangular resolution, wide field of view and multiwave (multicolor)spectral range. New large single mm/submm telescopes are in high demand, as well as their inclusion in the global Event Horizon Telescope (EHT) VLBI network. At the same time, there are no large mm/submm telescopes in Asia at allwhile appropriate sites exist and their appearance in Asia or Eurasia is long overdue.Kinetic inductance detectors (KID) are ideal for large-format array implementation, which will be necessary for future telescope development. Concept of multicolor subTHzKID-array MUSICAM demo camera and its instrumental testing is given. It allows us to perform some necessary steps toward the creation of the Eurasian SubMillimeter Telescopes (ESMT), which  concept and scientific tasks are presented as well.

## 1. Introduction

Presently, the range of possible scientific tasks for a large millimeter/submillimeter range telescope has grown significantly, from studies of star formation across the Universe to search for the Sunyaev–Zeldovich effect in galaxy clusters and primordial (cosmological) molecules in the early Universe for studies of the epoch of Dark Ages. Another intriguing task is the investigation of nature of dark matter and dark energy. New large single mm/submm telescopes are in high demand, as well as their inclusion in the global Event Horizon Telescope (EHT) VLBI network. At the same time, there are no large mm/submm telescopes in Asia at all, and their appearance in Asia or Eurasia is long overdue.

New challenges in submillimeter wave astronomy require instruments with a combination of high sensitivity and angular resolution, wide field of view and multiwave (multicolor) spectral range. One commonly used detector technology for submillimeter wavelengths proposed in 1994 is the transition edge sensor (TES), which is a cryogenic sensor based on the strongly temperature dependant resistance near the superconducting phase transition. To read the signal from a TES, a superconducting quantum interference device (SQUID) is coupled with the detector. TES have been used for the detection of submillimeter/millimeter radiation in many types of instruments; however, their complex fabrication process and readout method make them difficult to scale to larger arrays. Relatively new type of sensor is a kinetic inductance detector (KID; also MKID, where M stands for microwave) first developed at the Caltech and the Jet Propulsion Laboratory (JPL) in the early 2000s. Its action is based on change in the surface impedance of a superconductor due to increase of kinetic inductance caused by the incident photons. Kinetic inductance is a manifestation of the inertial mass of charge carrier in alternating electric field as an equivalent series inductance. In superconductors, kinetic inductance is in inverse proportion to the concentration of Cooper pairs. The incident photons break Cooper pairs, which yields an extra inductance. This change can be accurately measured by placing such a superconducting inductor in a lithographed resonator. A microwave probe signal is tuned to the resonant frequency of the resonator. Any photons absorbed in the inductor will imprint their signature as changes in phase and amplitude of the probe signal.

KIDs can be easily fabricated on a two- to three-layer wafer and frequency-domain multiplexed; all of the readout functions are performed by room-temperature electronics, with the exception of one cryogenic amplifier. KIDs are ideal for large-array implementation, which will be necessary for future telescope development. For many applications, MKIDs are the first choice, with much simpler fabrication, operation and lower cost than TES. The examples of TES design areSCUBA-2[1] and BICEP2[2], the MKID cameras are MUSIC[3] and NIKA2[4].

Both competing TES and KID technologies are now used for submm astronomy, but the latter definitely wins this competition when it comes to building large-format arrays for telescopes with large field of view. In particular, MKID technology will be used for the CCAT-prime telescope with a field of view of 1 square degree. The NIKA2 KID camera was designed for the 30 m IRAM telescope with 2900 pixels covering FOV=6.5′×6.5′in two wavelength ranges of 2.1 mm and 1.25 mm, however this is achieved by using two separate arrays, which implies complex matching optics of the telescope[4]. The multicolor MKID array camera MUSIC, designed for CSO, solves the problem by combining 2304 detectors for different ranges in one sensor covering field of view of 14′×14′[3].

The main goal of the proposed concept of MUSICAM demo camera and its instrumental testing is to perform some necessary steps toward the creation of the Eurasian SubMillimeter Telescopes (ESMT).

## 2. ESMT concept and atmospheric limitations

In Table 1, the largest mm/submm telescopes and antenna arrays, either currently operating or to come into operation until 2025 (the latter marked with *), are listed.

Table 1

| Facility | Location | Diameter, m | Altitude, m | Shortest wavelength, mm |
|---|---|---|---|---|
| LMT | Mexico (Sierra Negra) | 50 | 4600 | 0.85 |
| IRAM | Spain (Sierra Nevada) | 30 | 2850 | 0.9 |
| JCMT | Hawaii (Mauna Kea) | 15 | 4092 | 0.4 |
| APEX | Chile (Atacama) | 12 | 5058 | 0.23 |
| GLT* | Greenland (Summit, now in a test mode in Thule) | 12 | 3210 | 0.2 |
| NOEMA* | France (Alps, Plateau de Bure) | 15x12 | 2550 | 0.8 |
| ALMA | Chile (Atacama) | 12×54 + 7x12 | 5100 | 0.32 |
| SPT | South Pole | 10 | 2800 | 0.4 |
| ARO | Kitt Peak, Arizona, US | 12 | 1895 | 0.6 |
| SMT | Mount Graham, Arizona, US | 10 | 3185 | 0.3 |
| CCAT-prime* | Chile (Atacama) | 6 | 5616 | 0.2 |

Table 1 shows that there are no large mm/submm radio telescopes in Asia at all, while appropriate sites exist. Astronomical observations are usually carried out in transparency windows, i.e. at local minima of the absorption spectrum (Fig. 1). Previously, the joint research group of the IAP RAS and NSTU created the equipment and developed methods for studying the astroclimate in the subTHz range. They investigate spectral characteristics of atmospheric transparency in laboratory conditions. Since 2012, regular field studies of the astroclimate have been performed for searching for new sites suitable for subTHz observations in Eurasia [5] (Fig. 1).

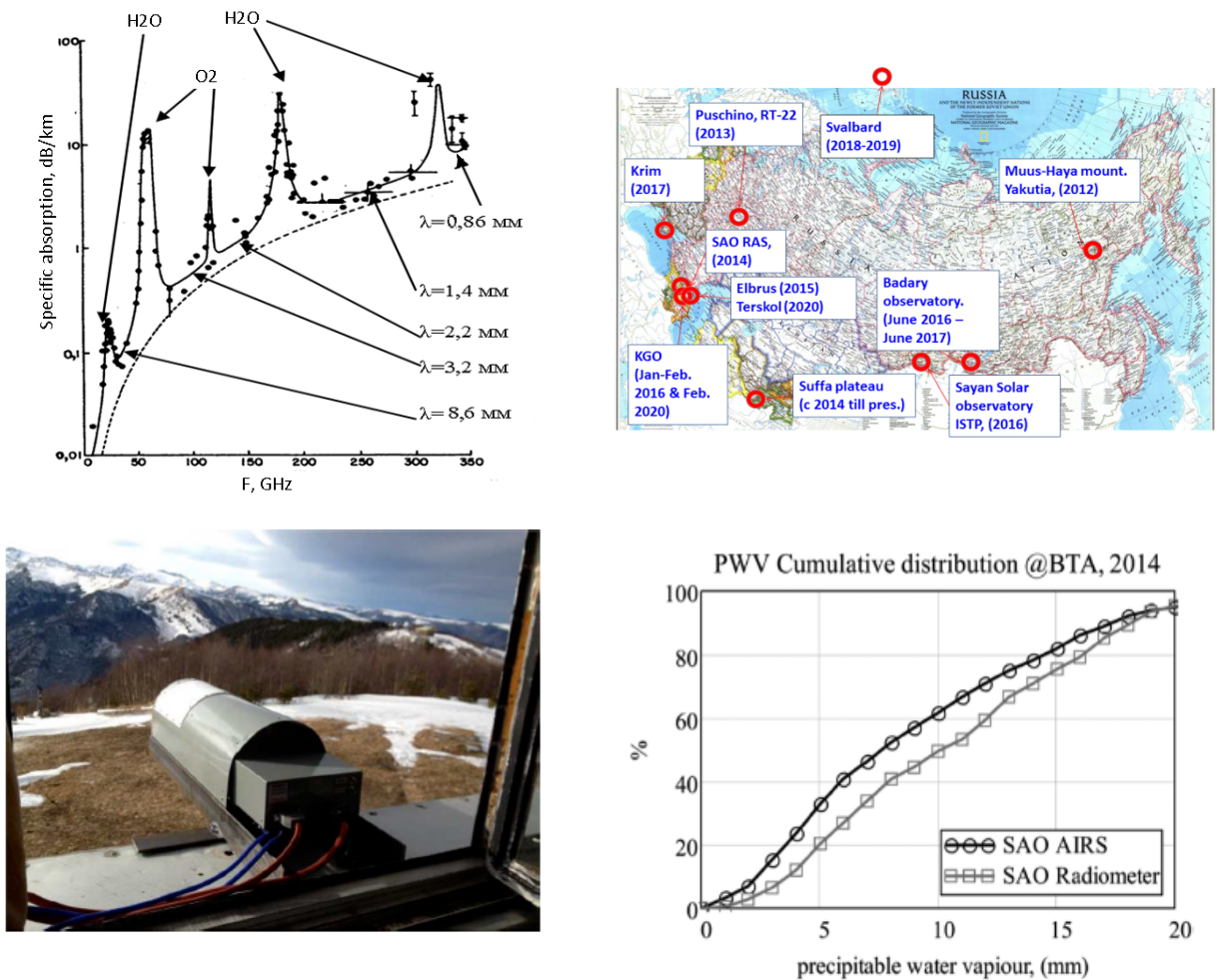

Fig. 1. Specific attenuation of radio waves in atmosphere with transparency windows at a pressure of 1 atm, a temperature of 20°C and a water vapor concentration of 7.5 g/m$^2$ (top left), sites tested for the astroclimate conditions (top right), tipping radiometer in BTA tower window (bottom left), PWV cumulative distribution for BTA in 2014 (bottom right)

The essence of the ESMT concept is in realizing the impossibility of the RT-70 mm/submm radio telescope in Suffa, Uzbekistan, and proposing, instead of completion of the old huge Soviet project and its subsequent modifications, the construction ofseveralsmallerbut effective mm/submm telescopes. The concept of the ESMTproject implies building three new mm/submm telescope of 21 m class located in Suffa, Uzbekistan (2500 m above sea level), Russia (above 3000 m a.s.l) and in Tibet, China (above 4000 m a.s.l.). The ESMT concept was developed by specialists in precise antennas, astroclimate research and subTHz matrix receivers from SAO RAS,IAP RAS and NAOC togetherwith EIE Group (www.eie.it).It was suggested to take the design of European ALMA antennas (12 m) as a starting point, increase their diameter to the maximum possible without fundamental changes in the design of the truss frame and panels, and supplement them with an "active surface". The well-developed, closed, thermostable high modulus carbon fiber reinforced polymer truss structure preserves its shape, while all its dimensions increase proportionally. Mechanical and thermal modeling performed inEIE Group has shown the possibility of such a scaling for the antenna diameter up to 20-21 m[6].The expected accuracy of the surface of the main mirror is not worse than 25 μm (RMS) for a diameter of 21 m and 15 μm (RMS) for a diameter of 15 m. The upper frequency band for such a telescope inthe best Tibet atmospheric conditionsis 1.5 THz (Ali2), 0.95 THz(Ali1), 0.85 THz (Yangbajing). Expected PWV and transmission for Yangbajing site in comparison with Atacama desertsites, Mauna Kea and Summit Campgiven in the Table 2 for 0.86 THz and 1.5 THz were taken from [7]. PWV data in round bracketsfor Ali1, Ali2,and Cerro Chajnantor were taken from [8]; these values obtained later seem to be up to 20% higher than values from [7]. The Table 2 shows that sites Cerro Chajnantor and Ali2 are quite equal;both thesesites arestudied and developed now for astronomy research.

Table 2

| Site | Altitude, m | PWV, mm | | Minimum transmission, % (λ=350μm) | | Minimum transmission,% (λ=200 μm) | |
|---|---|---|---|---|---|---|---|
| | | Time | | Time | | Time | |
| | | 10% | 25% | 10% | 55% | 10% | 25% |
| Yangbajing, Tibet, China(CCOSMA) | 4300 | 1.21 | 2.45 | 19 | 2 | 0 | 0 |
| Ali1, Tibet, China (AliCPT) | 5250 | (0.55) | (0.871) | <49 | <41 | <5 | <2 |
| Ali2, Tibet, China | 6100 | (0.25) | (0.459) | >65 | >58 | >20 | <10 |
| Chajnantor plateau, Chile (ALMA, APEX) | 5100 | 0.39 | 0.53(0.618) | 56 | 46 | 9 | 4 |
| Cerro Chajnantor, Chile(CCAT) | 5612 | 0.27 | 0.37(0.48) | 65 | 58 | 20 | 12 |
| Cerro Macon,Argentina | 5032 | 0.47 | 0.66 | 49 | 41 | 5 | 2 |
| Summit Camp, Greenland(GLT) | 3210 | 0.36 | 0.51 | 42 | 31 | 3 | 1 |
| Mauna Kea,Hawaii(JCMT) | 4100 | 0.62 | 0.91 | 40 | 27 | 3 | 1 |

The expected zenith Planck brightness temperatures $T_B$ at220 GHz for Ali1 and Ali2 are 30-34 K and 19-23 Kfor 75% of time[8].

## 3. Scientific tasks for ESMT

Observations withlarge millimeter/submillimeter telescopes are important for solving a number of topical astrophysical problems. These include, for ESMT, in particular, the following.

Studies of the star formation process. Despite the large number of works in this direction, many important aspects of this process remain unclear. This is especially true for the formation of massive stars. The peak of radiation from star-forming regions in spectral lines and in the continuum falls in the submillimeter range, which makes the observations of such objects in this range especially important. Studies of star-forming regions will be carried out both in our galaxy and in other galaxies.

Study of galaxy clusters based on observations of the Sunyaev–Zeldovich effect. This task is now especially urgent in connection with the successful start of the operation of the Spektr-RG space observatory, which records the X-ray radiation of these objects. The combination of observational data in the X-ray and radio ranges allows one to estimate both the physical characteristics of the clusters and some cosmological parameters.

Research of sources of gravitational waves and gamma-ray bursts. A burst of gravitational waves from merging black holes was first recorded several years ago. Since then, several similar events have been recorded, including such a burst from the merger of neutron stars. Till now, there is almost no identification of these events with known astronomical objects. To identify and study the properties of these sources, observations in various ranges, including those at submillimeter waves, are important. Gamma-ray bursts have been observed for a long time, but the mechanism of their generation is still unclear. Observations of the emission spectra of gamma-ray burst sources in different wavelength ranges are important for solving this problem.

Search for new chemical compounds in space, in particular, complexorganic molecules. This task is important, in particular, for the search for life in the Universe, which is one of the main challenges for the mankind.

Search for distortions of the microwave background spectrum due to primary molecules and energy release in the early Universe [9,10]. Apparently, this is the only way to get information about the so-called "Dark Ages" of the Universe — the era between the recombination of the primary plasma and the formation of the first stars and galaxies. Such information will make possible the understanding of the processes of formation of structures in the early Universe, from which the observed objects arose.

Search for $HeH^+$ and other primordial molecules such as LiH, $HD^+$, $H_2D^+$ in the Universe [10,11].

Studies of supermassive black holes in the centers of galaxies as part of the global interferometric network "Event Horizon Telescope" (EHT). Quite recently, the EHT collaboration presented the first results of observations of the "shadow" of a supermassive black hole in the M87 galaxy [12]. The extension of this network with submillimeter telescopes in Asia (where they are currently absent) will significantly increase its capabilities and allow obtaining better images of such objects. The angular resolution of the terrestrial network so far allows us to study practically only two such objects, in M87 and in the center of our Galaxy. In the future, with the advent of a space telescope of the millimeter range (the Millimetron project), the list of objects will be significantly expanded.

## 4. Concept of the MUSICAM demo

Just like the MUSIC Multiwavelength Sub/millimeter Inductance CAMera, the proposeddemo device (the MUSICAM-demo camera)will combine sensors for 4 wavelength ranges corresponding to water vapor transparency windows of 2.0 mm, 1.33 mm, 1.04 mm, and 0.86 mm (Fig. 1), with 5×5×4 = 100 pixels. Broadband superconducting phased-array slot-dipole antennas will be used to form beams, and lumped element on-chip bandpass filters willdefine spectral bands. To achieve physical temperature of 250 mK a compact dissolution cryostat will be used [13].

### 4.1. Design of camera wafer

As shown in Fig. 2,signal is first received by the phased-array antenna and then split into different frequency bands determinedby the BPF network. The output of each filter is fed into one LC resonator, which is coupled to thefeedline.

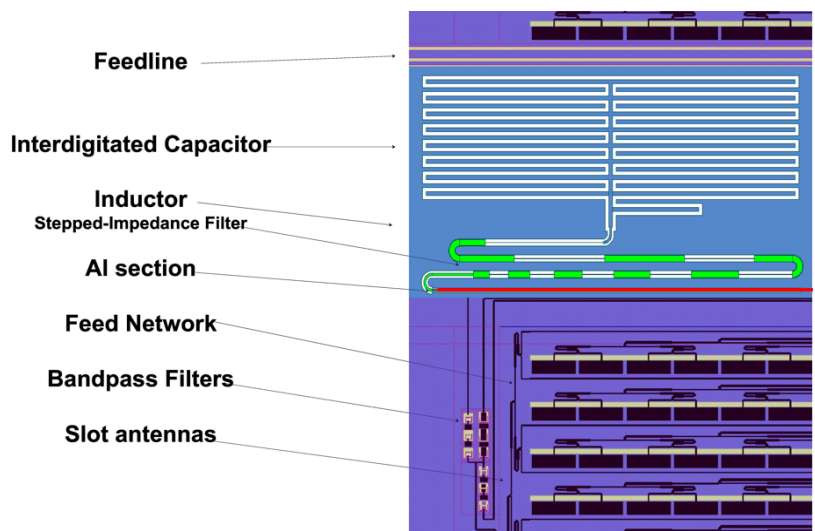

Fig.2. Mask design: layout of the major components on the wafer.

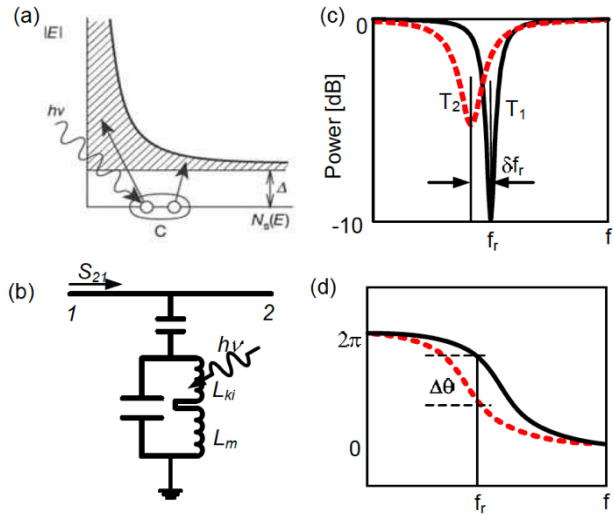

Fig.3. Schematic illustration of MKID operation (a) A photon breaks Cooper pairsand creates quasiparticles in a superconducting strip. (b) The increase in the quasiparticle densitychanges thesurface impedance. (c) The transmission through the resonant circuit exhibits a narrowdip at the resonance frequency $f_r$, which shifts when the surface impedance changes. (d) Themicrowave probe signal experiences a phase shift when $f_r$ changes.

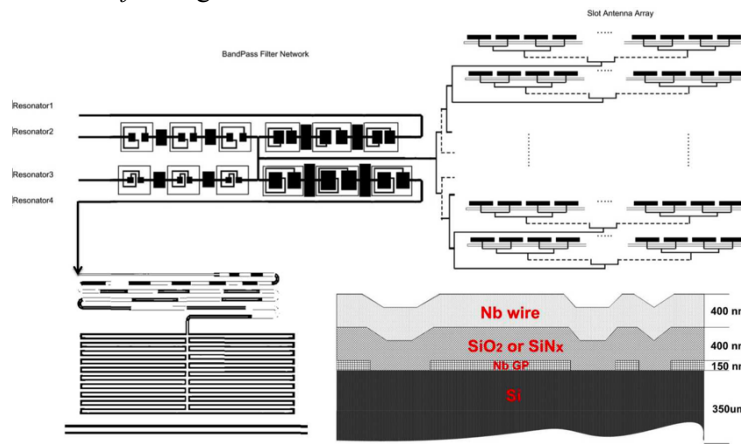

Fig.4. Scaled version of the antenna, BPF network, and KID layout. (Upper right) A phased-array antenna with a binary summing tree is used to capture the signal. (Upper left) Four bandpassfilters split the signal from the summing tree into four different frequency bands. (Lower left). The kinetic inductance detector layout. (Lower right) Substrate layers of the detector wafer.

As shown in Fig. 3, when the aluminum superconductor receives photons, the photons breakCooper pairs and create quasiparticles, which change the surface impedance of the aluminum superconductor. Because the aluminum section is part of the meander inductor of the LC resonator, thephotons cause the inductance of the LC resonator to change, thereby affecting its resonance properties including the phase and power (i.e., a measure of how the resonator couples to the microstripline). By monitoring the resonance properties, we can detect signals at submillimeter wavelengths. We call such an LC resonator a microwave KID.Overall, the KID-based instrument that we developed is an ideal choice for use in submillimeterinstruments because of the following features:1. The detectors are frequency-domain multiplexed with hundreds of detectors coupled throughone feedline.2. The process for fabricating a large detector array is relatively simple, consisting of two to fourlayers of a lithographic process on a silicon wafer.3. The readout system for this large detector array has been developed and demonstrated tosatisfy large-array requirements.These advantages not only make the KID-based instrument scalable to larger arrays but alsoallow for cost control on a per-pixel basis.

## 4.2. Readout for Kinetic Inductance Detectors

The readout system will perform frequency domain multiplexed real-time measurements of complex microwave transmission coefficient in order to monitor the instantaneous resonant frequency of superconducting microresonators and changes in their energy dissipation. Each readout unit similar to those used in the MUSIC device will be able to cover up to 550 MHz bandwidth and read 256 complex frequency channels simultaneously[14]. The digital electronics include the customized DAC, ADC, IF system and the FPGA based signal processing hardware developed by CASPER group.

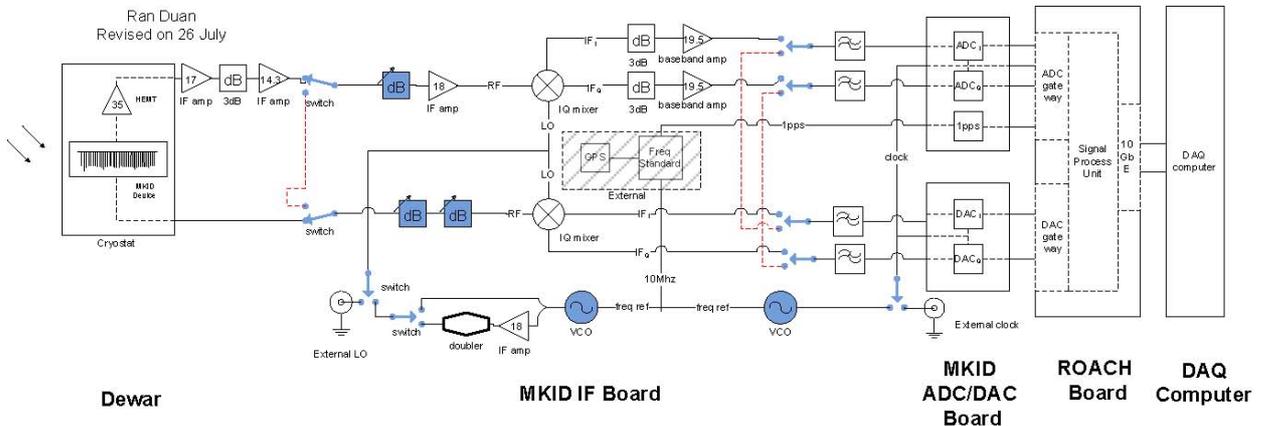

Fig.5. Readout system block diagram.

In general, the procedure for the KID readout is asfollows:

1. A drive tone or carrier tone (which generally ranges from approximately 100 MHz to a fewgigahertz) is sent through the transmission line on the device with which the detector iscoupled. The drive tone is then modulated by the detector as it responds to the power of theastronomical signal.

2. If the modulated carrier tone is at a frequency that is too high to be directly digitized, it isdown-converted using a mixer.

3. The received tone is digitized and processed. This step includes digitizing the analog signaland processing the digital signal using a field-programmable gate array (FPGA), a graphicsprocessing unit (GPU), a central processing unit (CPU), or some combination thereof. Thesignal from the detector generally requires real-time processing to capture the signal source,reduce the data rate, and extract useful information.

4. Auxiliary information such as a timestamp or information regarding the telescope is stored onthe data acquisition (DAQ) computer.

5. Frequently, the raw data must be subjected to additional computer-based processing stepsprior to use. These steps may include serialization of the data stream, noise reduction, trans-formation of the data format or units, calibration, or mapmaking.

## 4.3. FPGA Firmware

A flow chart that provides an overview of the firmware behavior is presented in Figure below.The lower diagram shows the DAC LUT and DAC gateway, which is the direction in which theFPGA plays back the LUT buffer and sends the signal to the DAC board. This pathway includesa serialization step to provide the DAC with four DAC clock cycles of data during each FPGAclock cycle. The upper diagram shows the signal receiving and processing chain on the FPGA. Onthe receiving side, the ADC outputs the data in parallelized (deserialized) format, i.e., four ADCclock cycles for each FPGA clock cycle. The corresponding data streams from the I and Q ADCsare added to obtain a complex number, and the four complex streams are then processed by fourparallel FFT cores (housed in two Biplex FFT blocks). Of the 216 bins output by the four parallelFFTs, we are interested only in the 192 that have carrier tones; therefore, the data stream passesthrough a frequency-bin selector, and then, the 192 data streams pass through a 192-channel FIR filter to decimate the data rate by a factor of 75, which makes the final data rate exactly 100 Hz.Before the data are sent out of the FPGA, we package the data stream into a transmission controlprotocol (TCP) packet with a timestamp.

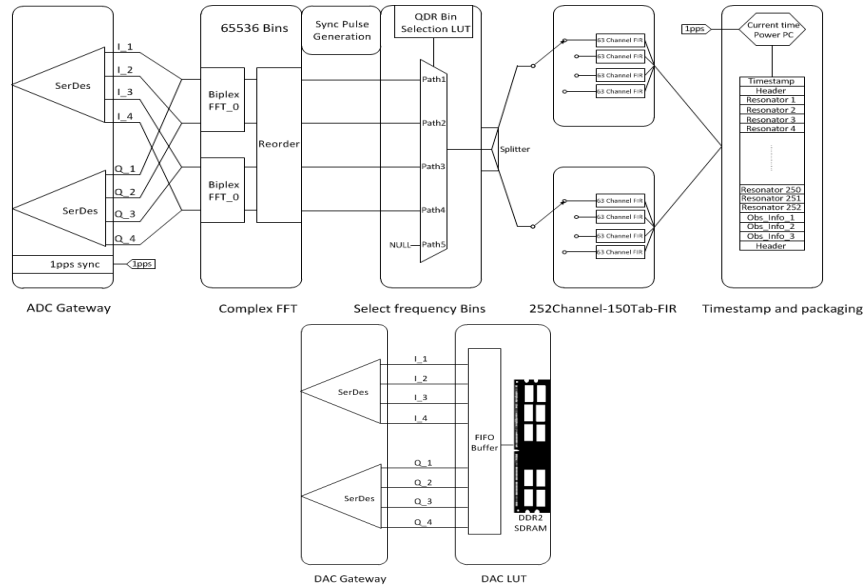

Fig.6. Flow chart of the FPGA firmware: (top) from the ADC input to the FPGA output and(bottom) lookup table and DAC.

### 4.4. Data Acquisition System and Instrument Test Operation

We have discussed both hardware and firmware. In addition to these components, we also need to develop a software package to fully automate the process of readout, hardware control, communication with the telescope, and computer-controlled data acquisition. We call this software package the DAQ system, and a block diagram of this system is presented in Figure below.

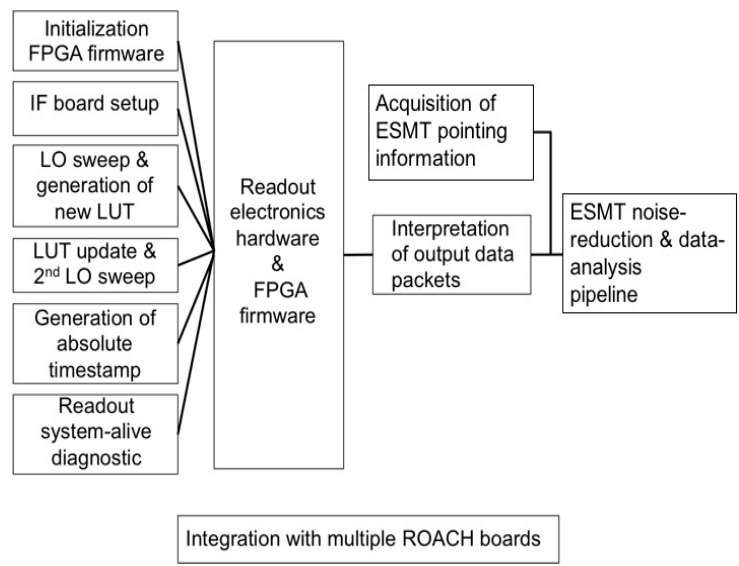

Fig. 7. Block diagram of the DAQ functions and processes.

Under clear sky conditions and PWV < 5, the expected sensitivity of the MUSICAM-demo camera in antenna temperature is 0.6 mK·$s^{1/2}$ at $\lambda = 2$ mm.

In a compact dissolution cryostat[13] the effect of deep cooling in a cooled ampoule is achieved by creating a circulation of helium-3 in a mixture of helium-3 and helium-4 due to pumping of helium-3 vapors from the mixture caused bytheir condensation in a chamber cooled by pumping out using a sorption pump of pure helium-3. Using this technology in dissolution cryostats is know-how of IPP RAS, which determines the unique features of such devices: compactness, complete autonomy, the absence of mechanical units operating at low temperatures, the possibility of 100% automation.

The MUSICAM-demo can be used for the continuum observations and some spectral astrophysical and cosmological tasks such as search for primordial molecules in the Universe. For spectral tasks, an external quasi-optical wide-aperture resonant metal mesh band path filter (BPF) may be used with a bandwidth of about 1% of central frequency of the detector's spectral range. Then subbands up to 550 MHz wide within the BPF bandwidth can be analyzed with an FFT spectrometer attached to the output of the MUSICAM-demo.

## 5. Instrumental tests of MUSICAM-demo camera

For instrumental tests of the MUSICAM-demo camera in all four bands Russian telescopes BTA (Nizhny Arkhyz, 2100 m) and Zeiss-2000 (Terskol, 3100 m) will be used. To use the MUSICAM-demo efficiently, tertiary and relay sub-THz quasi-optics must be developed in order to match the camera to a particular telescope. The optimal matching has to meet two requirements: (I) two celestial objects just resolved by the telescope optics must map on two adjacent pixels in the image plane; (II) elements of the receiver array must be fed appropriately, which in the case of the MUSICAM-demo receivers means that the output beam of the entire system lit by a parallel ray bundle must have an effective F# of 3.0. Given the telescope F# and its focal distance, as well as pixel spacingand overall dimensions ofthe sensor array, one can design the necessary matching system. The possible optical layouts for matching MUSICAM-demo with BTA and Zeiss-2000 and the focal spots in the image plane are presented in Fig. 8. The Airy disk calculated without taking apodization into account is shown as a circle in the center of the picture. The Strehl ratio for each of the beams is greater than 0.98 in both cases. Modeling shows that because of vignetting the matching optics for the BTA and Zeiss-2000 telescopes must be designed to provide certain oversampling, i.e. overlapping of the adjacent beams at level higher than -3 dB, which will allow for simultaneous testing of all the pixels at the expense of reducing the FOV of the system. Tests of the MUSICAM-demo camera in the wavelength ranges of 2 mm/1.3 mmand 1.08/0.86 mm are planned to be conducted at BTA and Zeiss-2000 respectively. The percentage of days suitable for qualitative observations at 2 mm and 1.3 mm with BTA is, according to AIRS data, more than 25 and 7, respectively. To assess the possibility of performing effective observations on Zeiss-2000 at wavelengths of 1.08 mm and 0.86 mm, a long-term study of the astroclimate in subTHz range at Terskol began. The first results of measurements of the optical depth at a wavelength of 2 mm in Terskol are encouraging. The latest astroclimate measurements of sites under study are given in [15].

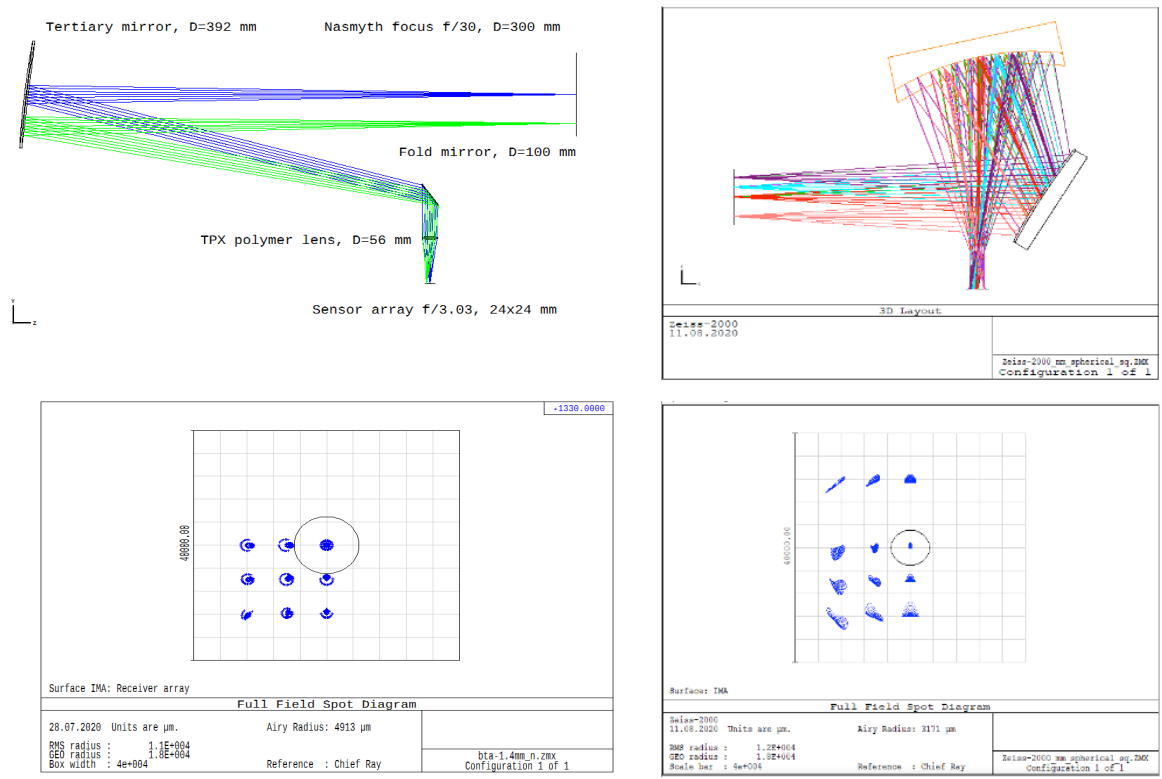

Fig. 8. The possible optical layouts for matching MUSICAM-demo with BTA and focal spots (left) and Zeiss-2000 (right)